\documentclass[aps,twocolumn]{revtex4}
\usepackage{epsf}
\usepackage{graphicx}

\begin{document}

\title{ Nos\`e-Hoover Dynamics in Quantum Phase Space
}

\author{Alessandro Sergi\footnote{E-mail: sergi@ukzn.ac.za}}
\affiliation{
School of Physics, University of KwaZulu-Natal, Pietermaritzburg Campus,
Private Bag X01 Scottsville, 3209 Pietermaritzburg, South Africa
}

\author{Francesco Petruccione\footnote{E-mail: petruccione@ukzn.ac.za}}
\affiliation{
School of Physics, Quantum Research Group,
University of KwaZulu-Natal, Westville Campus, Private Bag X54001,
Durban 4000, South Africa
}

\begin{abstract}
Thermal fluctuations in time-dependent quantum processes
are treated by a constant-temperature generalization of Wigner's 
formulation of quantum mechanics in phase space.
To this end, quantum Nos\`e-Hoover dynamics is defined
by generalizing the Moyal bracket.
Computational applications of the formalism, together with further
theoretical developments, are discussed.
\end{abstract}

\maketitle

Many (equivalent) ways to represent
quantum mechanics in phase space~\cite{psqm} have been developed since
the original paper of Wigner~\cite{wigner}.
Although the phase space is comprised
of the classical coordinates of the system $(r,p)$, positions and momenta
respectively, 
it is nevertheless quantum since the laws of motion and the statistical
constraints of the theory are designed so that
standard quantum averages can be reproduced.
From a formal point of view, we may note that, within all the different
formulations of quantum mechanics in phase space, Wigner's provides
perhaps the most simple one. Wigner's formulation of quantum mechanics
has been recently used to investigate quantum states of the electromagnetic 
field~\cite{atomlaserpulse}, quantum-to-classical correspondence
both in the electronic transport properties in nanowires~\cite{nanowires}
and in driven nonlinear nanomechanical resonators~\cite{resonator}.

Wigner's approach to quantum mechanics can be formulated by
means of the Moyal bracket~\cite{moyal}.
This provides an algebraic route to the formulation
of quantum mechanics in phase space~\cite{moyal,groenewold}
which has also been exploited
in what seems to be an emerging branch of quantum theory,
known as \emph{deformation quantization}~\cite{deformation}.
As a matter of fact, Moyal's bracket is obtained within the
deformation quantization approach upon \emph{deforming}
the well-known classical Poisson structure in order to introduce
a generalized bracket in phase space expressing quantum laws.
We note that a strategy similar to that used in deformation quantization,
\emph{i.e.}, the generalization of brackets, has been adopted
very recently to introduce a formalism to derive generalized equations of motion and
analyze the statistical mechanics of non-Hamiltonian systems both
in the classical~\cite{sergi-c} and in the quantum context~\cite{sweinberg,sergi-q}.
In particular,  it has been shown~\cite{qc-nose} how to formulate
constant-temperature dynamics for quantum-classical systems
so that a reduced number of classical bath degrees of freedom
can be used to simulate relaxation dynamics.
This required to combine the method of constant-temperature dynamics~\cite{nose}
(known as Nos\`e-Hoover dynamics) used in computer simulations of
classical systems with the quantum-classical dynamics~\cite{sergi-q}.
In classical molecular
dynamics the Nos\`e-Hoover equations~\cite{nose}, and their variants~\cite{nhc},  sample
the canonical ensemble very efficiently
and they can all be formulated as 
non-Hamiltonian phase space flows.

In many instances,
ranging from quantum phase transitions~\cite{subir} 
to lattice gauge theory simulations~\cite{rothe},
one is interested in calculating quantum averages in the canonical ensemble.
Hence, one could enquire whether,
within the Wigner formulation of quantum mechanics in phase space,
a non-Hamiltonian generalization of the Moyal bracket can be introduced
and used to investigate
the effect of thermal fluctuations in the applications
mentioned above~\cite{atomlaserpulse,nanowires,resonator}.

In this letter we introduce
constant-temperature Nos\`e-Hoover
dynamics~\cite{nose} in quantum phase space.
To this end, we generalize 
the Moyal bracket~\cite{moyal,deformation}
in a non-Hamiltonian fashion~\cite{sergi-c,sweinberg,sergi-q}.
Such a generalization allows one
to devise systematic approaches to the calculation of
quantum corrections to equilibrium or time-dependent properties
taking explicitly into account thermal fluctuations.

Given the von Neumann density matrix $\hat{\rho}$,
the quantum average of an arbitrary operator $\hat{\chi}$
can be calculated in phase space  
by introducing the Wigner transform of both the operator
and the density matrix:
$\chi_W(r,p)={\cal W}(\hat{\chi})$, 
$f_W(r,p)=(2\pi\hbar)^{-N}{\cal W}(\hat{\rho})$,
where ${\cal W}$ denotes the operator which realizes 
the Wigner transform~\cite{wigner},
$N$ is the number of degrees of freedom of the system,
and $(r,p)$ are the phase space coordinates.
The phase space quantity $f_W(r,p)$ is known as the Wigner function.
Then, the quantum average 
$\langle\hat{\chi}(t)\rangle={\rm Tr}(\hat{\chi}(t)\hat{\rho})$
can be calculated as
\begin{eqnarray}
\langle\hat{\chi}\rangle&=&
\int d^Nrd^Np~ \chi_W(r,p,t)f_W(r,p)\;.
\label{eq:ave}
\end{eqnarray}
The law of motion for an arbitrary operator $\hat{\chi}$ can be found by Wigner-transforming
the Heisenberg equation of motion:
\begin{eqnarray}
\partial_t\chi_W(r,p,t)=\frac{i}{\hbar}\left\{H_W,\chi_W(t)\right\}_M\;,
\label{eq:moyaleqofm}
\end{eqnarray}
where $\partial_t\equiv\partial/\partial t$, $H_W={\cal W}(\hat{H})$
is the Wigner equivalent of the Hamiltonian operator $\hat{H}$.
The right hand side of Eq.~(\ref{eq:moyaleqofm}) defines the Moyal bracket,
which for two arbitrary (quantum) phase space 
functions, $\chi_W^1$ and $\chi_W^2$,
can be written as
\begin{eqnarray}
\left\{\chi_W^1,\chi_W^2\right\}_M
&\equiv& \chi_W^1\exp\left[\frac{i\hbar}{2}\overleftarrow{\partial}_i{\cal B}_{ij}^c
\overrightarrow{\partial}_j\right]\chi_W^2
\nonumber\\
&-&
\chi_W^2\exp\left[\frac{i\hbar}{2}\overleftarrow{\partial}_i{\cal B}_{ij}^c
\overrightarrow{\partial}_j\right]\chi_W^1\;,
\label{eq:moyalbracket}
\end{eqnarray}
where we have indicated the phase space point
through the compact notation $x=(x_1,x_2)=(r,p)$, and
$\partial_i$ stands for $\partial/\partial x_i$.
Note that we are using Einstein's
convention of summation over repeated indices, $i,j$ which go from $1$ to $2N$,
and we have
introduced the antisymmetric matrix
\begin{eqnarray}
\mbox{\boldmath$\cal B$}^c
&=&\left[\begin{array}{cc} {\bf 0} & {\bf 1} \\ -{\bf 1} & {\bf 0}\end{array}\right]\;.
\end{eqnarray}
Equation~(\ref{eq:moyalbracket}), which recasts the Moyal bracket in matrix form,
is central to the formalism that will be presented in the following.
Defining the operator
\begin{eqnarray}
\overrightarrow{\cal M}
&\equiv&
H_W\left(e^{\frac{i\hbar}{2}\overleftarrow{\partial_i}{\cal B}_{ij}^c
\overrightarrow{\partial}_j}
-
e^{-\frac{i\hbar}{2}\overleftarrow{\partial_i}{\cal B}_{ij}^c
\overrightarrow{\partial}_j}\right)\;,
\label{eq:moperator}
\end{eqnarray}
the propagator and the time evolution of observables can be written as
\begin{eqnarray}
\chi_W(r,p,t)=\exp\left[\frac{it}{\hbar}\overrightarrow{\cal M}\right]
\chi_W(r,p)\;.
\label{eq:dyna}
\end{eqnarray}
Together Equations~(\ref{eq:ave}) and~(\ref{eq:dyna}) specify how to calculate
quantum averages in phase space.
By means of repeated partial integrations, the time dependence in~(\ref{eq:ave})
can be transferred from the observable $\chi_W$ to the Wigner function $f_W$,
eventually recovering Wigner's equation of motion for the latter~\cite{wigner}.
For standard quantum dynamics, this amounts to substitute $f_W$ for $\chi_W$
in Eq.~(\ref{eq:dyna}) and having a minus sign in the propagator
($\exp\left[-\frac{it}{\hbar}\overrightarrow{\cal M}\right]$).
In a nutshell, this expresses the essence of the Wigner-Moyal formulation of quantum mechanics.

The matrix form of the Moyal bracket, introduced in Eq.~(\ref{eq:moyalbracket}), 
provides a mathematical structure that
can be naturally generalized for defining a 
Nos\`e-Moyal bracket with a non-Hamiltonian 
character. 
In order to introduce such a Nos\`e-Moyal bracket,
we extend the quantum phase space by adding two  Nos\`e variables
$\eta,p_{\eta}$, with fictitious mass $m_{\eta}$,
as it is done in the classical Nos\`e-Hoover molecular dynamics
approach. Hence, the phase space point coordinates are given by
$x=(r,\eta,p,p_{\eta})$. Correspondingly, we introduce 
a quantum extended Nos\`e Hamiltonian
\begin{eqnarray}
H_W^{\rm N}(x)=H_W(r,p)+\frac{p_{\eta}^2}{2m_{\eta}}+gk_BT\eta\;,
\end{eqnarray} 
where $g$ is the number of degrees of freedom, $k_B$ is Boltzmann's constant,
and $T$ is the fixed temperature of the canonical ensemble.
In the extended phase space, we can now generalize
the symplectic matrix
$\mbox{\boldmath$\cal B$}^c$ to
the antisymmetric tensor~\cite{sergi-c,sergi-q,qc-nose}
\begin{equation}
\mbox{\boldmath$\cal B$}^{\rm N}
=\left[\begin{array}{cccc} 0 & 0 & 1 & 0\\ 0 & 0 & 0 & 1\\
-1 & 0 & 0 & -p \\ 0 & -1 & p & 0\end{array}\right]\;,
\end{equation}
and re-define the Moyal bracket  in a way suitable
for the derivation of Nos\`e-Hoover dynamics.
In analogy to the definition~(\ref{eq:moperator}), we can use the antisymmetric tensor
$\mbox{\boldmath$\cal B$}^{\rm N}$ to introduce 
an operator in the extended quantum phase space
\begin{eqnarray}
\overrightarrow{\cal M}^{\rm N}
&\equiv&
H_W^{\rm N}\left(e^{\frac{i\hbar}{2}\overleftarrow{\partial_i}
{\cal B}_{ij}^{\rm N}
\overrightarrow{\partial}_j}
-
e^{-\frac{i\hbar}{2}\overleftarrow{\partial_i}{\cal B}_{ij}^{\rm N}
\overrightarrow{\partial}_j}\right)\;.
\label{eq:Mnose}
\end{eqnarray}
Equation~(\ref{eq:Mnose}) represents  the fundamental 
Nos\`e-Moyal operator for achieving Nos\`e-Hoover dynamics in quantum phase space.
The  time-propagation of the observable can be re-defined
by using the Nos\`e-Moyal operator $\overrightarrow{\cal M}^{\rm N}$ in place of $\overrightarrow{\cal M}$
in Eq.~(\ref{eq:dyna}).
Again, by means of repeated partial integrations, the time dependence
can be transferred from $\chi_W$ to $f_W$. However, in the extended phase space one finds
\begin{equation}
f_W(x,t)=\exp\left[-\frac{it}{\hbar}\overrightarrow{\cal M}^{{\rm N},\dag}
\right]f_W(x)\;,
\label{eq:qnosewigfun}
\end{equation}
with
\begin{eqnarray}
\overrightarrow{\cal M}^{{\rm N},\dag}
&=&
H_W^{\rm N}\left\{
\exp\left[\frac{i\hbar}{2}\left(
\overleftarrow{\partial}_i{\cal B}_{ij}^{\rm N}
\overrightarrow{\partial}_j
+\overleftarrow{\partial}_i\left(\partial_j{\cal B}_{ij}^{\rm N}\right)
\right)\right]\right.
\nonumber\\
&-&
\left.\exp\left[-\frac{i\hbar}{2}\left(
\overleftarrow{\partial}_i{\cal B}_{ij}^{\rm N}
\overrightarrow{\partial}_j
+\overleftarrow{\partial}_i\left(\partial_j{\cal B}_{ij}^{\rm N}\right)
\right)\right]\right\}
\;.\nonumber\\
\end{eqnarray}
This means that, under Nos\`e dynamics, the stationary Wigner function,
$f_{W,e}$, obeys the following equation
\begin{eqnarray}
-iL^{\rm N}f_{W,e}&-&\kappa^{\rm N} f_{W,e}
=\sum_{n=3,5,7,\ldots}\frac{1}{n!}\left(\frac{i\hbar}{2}\right)^{n-1}
\nonumber\\
&\times&H_W^{\rm N}\left[\overleftarrow{\partial}_i{\cal B}_{ij}^{\rm N}
\overrightarrow{\partial}_j+\overleftarrow{\partial}_i
\left(\partial_j{\cal B}_{ij}^{\rm N}\right)\right]^nf_{W,e}\;,
\nonumber\\
\label{eq:qnosestat}
\end{eqnarray}
where
\begin{eqnarray}
\kappa^{\rm N}&=& \left(\partial_j{\cal B}_{ij}^{\rm N}\right)
\partial_jH_W^{\rm N}=Np_{\eta}/m_{\eta}
\\
iL^{\rm N}&=&{\cal B}_{ij}^{\rm N}\left(\partial_j H_W^{\rm N}\right)
\overrightarrow{\partial}_i
\label{eq:fWnosestat}
\end{eqnarray}
are the Nos\`e phase space compressibility and the Nos\`e-Liouville
operators, respectively.
Equations~(\ref{eq:qnosewigfun}) and~(\ref{eq:qnosestat})
contain the full quantum corrections to the dynamics of all
the variables in the extended phase space.
They define Nos\`e-Hoover dynamics in a completely quantum fashion.
One of the main interests of such a generalization lies
in the fact that, as in the classical case, just a pair of additional
variables, namely the Nos\`e coordinates $(\eta,p_{\eta})$,
allows to represent thermal fluctuations (and therefore
the process of relaxation toward thermodynamical equilibrium)
of a quantum system
by introducing a suitable non-Hamiltonian dynamics 
in the extended phase space.
This is to be confronted with a standard Hamiltonian formalism
that would require coupling the physical coordinates, $(r,p)$,
to a bath composed by an infinite number of degrees of freedom. 

Once a stationary expression of the Wigner function 
has been found, equilibrium quantum averages that include thermal fluctuations
in the dynamical evolution (and not just in the initial conditions) can be calculated
by propagating observables according to
\begin{eqnarray}
\partial_t\chi_W&=&-\frac{2}{\hbar}H_W\sin\left[\frac{\hbar}{2}\overleftarrow{\partial}_i{\cal B}_{ij}^{\rm N}\overrightarrow{\partial}_j\right]\chi_W\;.
\label{eq:fullqnhdyna}
\end{eqnarray}
Equation~(\ref{eq:fullqnhdyna}) shows how to propagate a dynamical variable 
undergoing Nos\`e-Wigner time evolution in quantum phase space
taking into account quantum effects also in the dynamics of the
Nos\`e coordinates. In order to make clear the characteristics
 of the quantum corrections, one can expand in a Taylor series the sine 
in the right hand side of Eq.~(\ref{eq:fullqnhdyna}).
Upon realizing that 
$\partial^2 H_W^{\rm N}/\partial \eta^2=
\partial^3 H_W^{\rm N}/\partial p^3
=\partial^3 H_W^{\rm N}/\partial p_{\eta}^3=0$
and that all mixed derivatives
$\partial^2 H_W^{\rm N}/\partial x_i\partial x_j=0$,
for $i\neq j$, Eq.~(\ref{eq:fullqnhdyna}) can be rewritten
as
\begin{eqnarray}
\partial_t\chi_W&=&iL^{\rm N}\chi_W\nonumber\\
&-&\sum_{n=3,5,7,\ldots}\frac{1}{n!}
\left(\frac{i\hbar}{2}\right)^{n-1}
\left\{
H_W^{\rm N}
 \frac{\overleftarrow{\partial}^n}{\partial r^n}
 \frac{\overrightarrow{\partial}^n}{\partial p^n}
\chi_W\right.
\nonumber\\
&+&\left.H_W^{\rm N}\left[
 -\frac{\overleftarrow{\partial}}{\partial p}
\left(
 \frac{\overrightarrow{\partial}}{\partial r}
 -
 p\frac{\overrightarrow{\partial}}{\partial p_{\eta}}
\right)\right.\right.
\nonumber\\
&+&\left.\left.\frac{\overleftarrow{\partial}}{\partial p_{\eta}}
\left(-\overrightarrow{\frac{\partial}{\partial\eta}}
+
 p\frac{\overrightarrow{\partial}}{\partial p}\right)
 \right]^n\chi_W\right\}\;.
\label{eq:fullqnhdyna2}
\end{eqnarray}


We remark that, since thermal effects are described 
within the dynamics, one can also study relaxation processes
by using a non-equilibrium Wigner function and evolving the
dynamical variable according to Eq.~(\ref{eq:fullqnhdyna}):
the dissipation will be forced by the non-Hamiltonian coupling to the Nos\`e variables,
which mymic the thermodynamic of a thermal bath.

The analysis of the stationary Nos\`e-Hoover distribution in quantum phase space
can be simplified by performing 
a classical approximation on the dynamics of the Nos\`e variables alone
and keeping all quantum corrections in the dynamics of the 
coordinates of the physical system $(r,p)$.
In fact, we remind that, when Nos\`e-Hoover dynamics is implemented
within molecular dynamics computer simulations, a mass
$m_{\eta}\approx N\times m$ is used in order to achieve a weak coupling
to the Nos\`e ``bath''. In such a way, while the temperature of the system
is controlled, the equilibrium dynamical properties
of the physical coordinates $(r,p)$ are not significantly modified.
In the present context, a small parameter
$\mu=\sqrt{m/m_{\eta}}$ is  naturally found in the theory so that
one is allowed to perform a classical limit on the Nos\`e
coordinates $(\eta,p_{\eta})$. In this way, a quantum-classical
description, along the lines described in~\cite{kapracicco},
naturally arises.
Therefore, within a quantum-classical approach, one could disregard all the quantum corrections 
on the evolution of the Nos\`e variables in the left hand side
of Eq.~(\ref{eq:qnosestat}).
Upon assuming a standard form for the
Hamiltonian of the physical degrees of freedom $H_W(r,p)=(p^2/2m)+V(r)$,
the Nos\`e-Hoover equation for the stationary Wigner distribution function
becomes
\begin{eqnarray}
\sum_{n=3,5,7,\ldots}&&\frac{1}{n!}\left(\frac{i\hbar}{2}\right)^{n-1}
V(r)\left[\overleftarrow{\partial}_r\cdot
\overrightarrow{\partial}_p
\right]^nf_{W,e}\nonumber\\
&=&
-(iL_{\rm N}-\kappa^{\rm N})f_{W,e}\;,
\label{eq:qcnhwignerstat}
\end{eqnarray}
where $\partial_r=\partial/\partial r$ and $\partial_p=\partial/\partial_p$.
Moreover, in order to calculate quantum averages of functions of $(r,p)$ 
alone we need to calculate the average of Eq.~(\ref{eq:fWnosestat})
over the now classical Nos\`e variables. This turns out to be identical to
the equations first proposed by Wigner~\cite{wigner},
who showed how to obtain
quantum corrections in the canonical ensemble in terms of an expansion 
of his distribution function
in even powers of $\hbar$~\cite{wigner}:
\begin{equation}
f_{W,e}= \sum_{n=0}^{\infty}f_{W,e}^{(n)}\;.
\label{eq:wign}
\end{equation}
The order zero solution, $f_W^{(0)}$, to Eq.~(\ref{eq:qcnhwignerstat}) 
is~\cite{sergi-c}
\begin{equation}
f_W^{(0)}\propto\delta\left(H_W^{\rm N}\right)e^{-\int \kappa^{\rm N}dt}
\;.\label{eq:f_w0}
\end{equation}
When averaging over $\eta,p_{\eta}$, $f_W^{(0)}$
becomes the Boltzmann's weight for the $(r,p)$ variables.
Moreover, averages of odd functions of the Nos\`e variables
over $f_W^{(0)}$ are zero.
Hence, the analysis of the correction terms, under the approximations
of no quantum effects on the Nos\`e variables, can proceed as originally
shown by Wigner~\cite{wigner}.
The higher order correction terms
will all contain the zero-order term:
$f_{W,e}^{(n)}=\hbar^nf_{W,e}^{(0)}\tilde{f}_{W,e}^{(n)}$, with $n\ge 2$.
In principle, such correction terms could be used in order to calculate averages
over an ensemble of dynamical trajectories including thermal fluctuations
by means of Nos\`e-Hoover dynamics as defined through the operator
$\overrightarrow{\cal M}^{\rm N}$ introduced in Eq.~(\ref{eq:Mnose}).

Previous attempts to formulate quantum Nos\`e-Hoover dynamics~\cite{grillitosatti}
run into the problem of the nonergodicity of the dynamics.
However, in the present case,
exploiting the antisymmetric matrix structure of Moyal bracket,
introduced Eqs.~(\ref{eq:moperator}) and~(\ref{eq:Mnose}),
one can easily
formulate more sophisticated thermostatting methods, such as
Nos\`e-Hoover chain dynamics~\cite{nhc},
in order to, at least, tame eventual ergodicity problems with dynamics
in quantum phase space.

In addition to the Nos\'e-Hoover dynamics in quantum phase space, 
we would also like to discuss
some interesting theoretical developments which arise
from the matrix structure of the Moyal bracket,
which has been introduced in Eqs.~(\ref{eq:moperator}) and~(\ref{eq:Mnose}).
Such a matrix structure allows suggests a connection between
the approach of deformation quantization and
a formalism proposed by Weinberg in order to address non linear effects 
in quantum mechanics~\cite{weinberg}.
Weinberg's formalism has been slightly generalized and 
expressed in matrix form in~\cite{sweinberg}.
Having introduced a generalized matrix structure for the Moyal bracket,
Weinberg's non-linear formalism can be re-cast within a deformation quantization
approach: one just needs to define more general Moyal brackets by means of
an antisymmetric tensor 
$\mbox{\boldmath$\cal B$}=\mbox{\boldmath$\cal B$}[f_W(x)]$,
which can be a functional of the Wigner distribution function itself.
To the authors's knowledge such mathematical structures have not been
been studied so far.
In such a way, one has at disposal a deformation quantization formalism
for addressing eventual non-linear phenomena in quantum mechanics.

From a practical point of view, the main
justification for the continuous attempts to modify or generalize the formalism
of quantum mechanics lies in the fact
that the devolopment of a general method to simulate quantum dynamics
still remains an open problem facing the physics community.
The availability of such a method, of course, defines
the very possibility of predicting and calculating quantities of interest
in many physico-chemical processes in condensed matter systems
(including biological ones).
Typically, one desires to formulate quantum properties in a framework
as close as possible to the classical one so that computational methods,
which are known to be efficient in classical mechanics, can be adapted
to study quantum time-dependent processes. 
For example, the path integral approach to quantum mechanics
provides an isomorphism between quantum and classical systems
which precisely permits to devise such computational schemes.
However, the above mentioned isomorphism cannot be easily exploited,
to address quantum-time dependent properties in an efficient way.
Other routes, such as the one we have sketched here, must be searched. 
As a matter of fact, there are various computational schemes that,
exploiting the theory of stochastic
processes~\cite{petruccione}, attempt to simulate either full quantum dynamics~\cite{drummond}
or quantum-classical dynamics in phase space~\cite{kapral}.
Deformation quantization is one of such routes and
we believe that, by generalizing the Moyal bracket, 
we have taken a first step toward the pursue of a mathematical language
for developing novel computational approaches.
Typically, the introduction of the Nos\`e-Moyal bracket
in quantum phase space, and the corresponding possibility
of treating thermal fluctuations through non-Hamiltonian dynamics,
opens a new route for treating open quantum systems~\cite{petruccione}.

\vspace{0.3cm}
\noindent
{\bf Acknowledgement}

\noindent
This work is based upon research supported by the South African
Research Chair Initiative of the Department of Science and Technology 
and National Research Foundation.



\begin{thebibliography}{99}

\bibitem{psqm}
M. Hillery, R. F. O'Connell, M. O. and Scully, E. P. Wigner,
Phys. Rep. {\bf 106}, 121 (1984);
H.-W. Lee, Phys. Rep. {\bf 259}, 147 (1995).

\bibitem{wigner}
E. Wigner, Phys. Rev. {\bf 40}, 749 (1932).


\bibitem{atomlaserpulse}
K. L. Moore, S. Gupta, K. W. Murch, and D. M. Stamper-Kurn,
Phys. Rev. Lett. {\bf 97}, 244101 (2006).

\bibitem{nanowires}
J. Feist, A. B\"aker, R. Ketzmerick, S. Rotter,
B. Huckestein, and J. Burgd\"orfer,
Phys. Rev. Lett. {\bf 97}, 150406 (2006).

\bibitem{resonator}
I. Katza, A. Retzker, R. Straub, and R. Lifshitz,
Phys. Rev. Lett. {\bf 99}, 071301 (2007).

\bibitem{moyal}
J. E. Moyal, Proc. Cambridge Philos. Soc. {\bf 45}, 99 (1949).

\bibitem{groenewold}
H. J. Groenewold, Physica {\bf 12}, 405 (1946).

\bibitem{deformation}
F. Bayen, M. Flato, C. Fronsdal, A. Lichnerowicz, and D. Sternheimer,
Ann. Phys. {\bf 111}, 61 (1978); Ann. Phys. {\bf 111}, 111 (1978).



\bibitem{sergi-c} A. Sergi and P. V. Giaquinta, 
J. Stat. Mech.: Theory and Experiment {\bf 02}, P02013 (2007);
A. Sergi, 
Atti Accad. Pelorit. Pericol. Cl. Sci. Fis. Mat. Nat.{\bf 87}, c1a0501003 (2005);
Phys. Rev. E {\bf 72}, 031104 (2005).
Phys. Rev. E {\bf 69}, 021109 (2004);
Phys. Rev. E {\bf 67}, 021101 (2003);
A. Sergi and M. Ferrario, 
Phys. Rev. E {\bf 64}, 056125 (2001).

\bibitem{sweinberg}
A. Sergi,
J. Chem. Phys. {\bf 126}, 074109 (2007);

\bibitem{sergi-q} 
A. Sergi, 
J. Chem. Phys. {\bf 124}, 024110 (2006);
Phys. Rev. E {\bf 72}, 066125 (2005).


\bibitem{qc-nose}
A. Sergi,
J. Phys. A {\bf 40}, (2007); 

\bibitem{nose}
S. Nos\`e, Mol. Phys. {\bf 52}, 255 (1984);
W. G. Hoover, Phys. Rev. A {\bf 31}, 1695 (1985);
S. Nos\`e,
Prog. Theor. Phys. {\bf 103}, 1 (1991).

\bibitem{nhc}
G. J. Martyna, M. L. Klein, and M. Tuckerman, J. Chem. Phys. {\bf 92}, 2635 (1992).

\bibitem{subir}
S. Sachdev, Quantum Phase Transitions (Cambridge University Press, Cambridge, 1999).

\bibitem{rothe}
H. J. Rothe, Lattice Gauge Theories: An Introduction.
(World Scientific, Singapore, 2005).

\bibitem{kapracicco}
R. Kapral and G. Ciccotti,
J. Chem. Phys. {\bf 110}, 8919 (1999).

\bibitem{grillitosatti}
M. Grilli and E. Tosatti,
Phys. Rev. Lett. {\bf 62}, 2889 (1989).

\bibitem{weinberg}
S. Weinberg, Phys. Rev. Lett. {\bf 62}, 485 (1989).


\bibitem{petruccione}
H.-P. Breuer and F. Petruccione, The theory of open quantum systems
(Oxford University Press, Oxford, 2002).

\bibitem{drummond}
P. Deuar and P. D. Drummond, Phys. Rev. Lett. {\bf 98}, 120402 (2007).

\bibitem{kapral}
A. Sergi and R. Kapral,
J. Chem. Phys. {\bf 121}, 7565 (2004).

\end{thebibliography}
\end{document}